\documentclass[10pt,oneside,a4paper]{article}

\usepackage{amssymb}
\usepackage{amsmath}
\usepackage{amsthm}
\usepackage{color}
\usepackage{graphicx}
\usepackage{wrapfig}
\usepackage{float}
\usepackage{subcaption}
\usepackage{rotating}
\usepackage{cite}
\usepackage{tikz}

\newcounter{thethm}[section]
\newtheorem{bigthm}{Theorem}[thethm]
\newtheorem{bigcor}[thethm]{Corollary}
\newtheorem{thm}{Theorem}[section]

\newtheorem{prop}[thm]{Proposition}
\newtheorem{lem}[thm]{Lemma}

\theoremstyle{remark}
\newtheorem{rem}[thm]{Remark}

\theoremstyle{definition}
\newtheorem{defi}[thm]{Definition}

\def\RR{\mathbb{R}}

\makeatletter
\newcommand{\labitem}[2]{%
\def\@itemlabel{\textbf{#1}}
\item
\def\@currentlabel{#1}\label{#2}}
\makeatother

\newcommand{\comment}[1]{}
\let\oldmarginpar\marginpar
\renewcommand\marginpar[1]{\-\oldmarginpar[\raggedleft\footnotesize #1]
{\raggedright\footnotesize #1}}
\author{
Jordi-Llu\'is Figueras
\thanks
{
Department of Mathematics, Uppsala University, 
Box 480, 75106 Uppsala (Sweden). {\tt figueras@math.uu.se}.
}
\and 
Dan Str\"angberg
\thanks
{
Department of Mathematics, Uppsala University, 
Box 480, 75106 Uppsala (Sweden). {\tt strangberg@math.uu.se}.
}
}

\title{
Non-Smooth Bifurcations of Uniformly Hyperbolic Invariant Manifolds
in Skew Product Systems: Rigorous Results}

\begin{document}
	
\maketitle

\begin{abstract}
In this paper we study the anti-integrable limit scenario of skew-product
systems. We consider a generalization of such systems based on the
Frenkel-Kontorova model, and prove the existence of orbits with any fibered
rotation number in systems of both one and two degrees of freedom. In
particular, our results also apply to two dimensional maps with degenerate
potentials (vanishing second derivative), so extending the results of existence
of Cantori for more general twist maps. 

We also prove that under certain mild regularity conditions on the potential
the structure of the orbits is of Cantor type. From our results we deduce the
existence of the non-smooth folding bifurcation (conjectured by Figueras-Haro,
\textit{Different scenarios for hyperbolicity breakdown in quasiperiodic area
preserving twist maps}, Chaos:25 (2015)). 

Lastly we present a pair of results which are useful in determining if a
potential satisfies the regularity conditions required for the Cantor sets of
orbits to exist and are also of independent interest.
\end{abstract}

\section{Skew product systems and their invariant sets}
\label{section: introduction}

Given a compact manifold $\mathcal M$, a skew product system is a map $(h,
F):\mathcal M\times \mathbb R^n\rightarrow \mathcal M \times \mathbb R^n$ with
$h(\theta, x)= h(\theta, y)$ for any $\theta\in\mathcal M$ and $x,y\in\mathbb
R^n$. Of interest are the ones with quasiperiodic dynamics on $\mathcal M =
\mathbb T^d$, $h(\theta)=\theta+\omega$, with $\omega\in\mathbb R^d$ being a
totally irrational vector (if $m\cdot \omega=0$ and $m\in\mathbb Z^d$ then
$m\equiv 0$). Robust smooth invariant manifolds are
\textit{fiberwise hyperbolic}: the vector bundle $\mathcal M\times \mathbb R^n$
decomposes in an invariant continuous Whitney sum $E^s\oplus E^u$ such that
there exists contants $C > 0$ and $0 < \lambda < 1$ satifying 
\begin{itemize}
\item $(\theta, v)\in E^s$ implies that 
$\|\pi_{\mathbb R^n}(h, D_2F)^k(\theta, v)\|\leq C\lambda^k\|v\|$ for $k\geq 0$;
\item $(\theta, v)\in E^u$ implies that 
$\|\pi_{\mathbb R^n}(h, D_2F)^k(\theta, v)\|\leq C\lambda^k\|v\|$ for $k\leq 0$.
\end{itemize}
A Fiberwise Hyperbolic Invariant Manifold (FHIM) satisfies that it is the graph
of a continuous function $K:\mathcal M\rightarrow \mathbb R^n$, $F(\theta,
K(\theta))=K(h(\theta))$, and persists under perturbations of $F$. See
\cite{HdLl06a, Figueras_Haro_CAP, Figueras_Haro_2015}.

The question of possible bifurcation scenarios of FHIM is of great interest. A
paradigm of this is the creation of Strange Non-Chaotic Attractors \cite{
Kaneko84, GOPY84, BezO96, Keller96, NK96, SFKP96, OF00, PrasadNR01,
Dattaetaltri04, HaroSimo05, HaroP06, SNA_book, GlendinningFPS00, OsingaWGF01,
KimLim04, AKdLL07, Jager_07, AlsedaMisiurewicz08, Bjerklov09, Jager09,
Jager09_memoirs}. These are one-parametric skew products on $\mathbb T\times
\mathbb R$ satisfying that a smooth attracting invariant curve bifurcates to an
only measurable attracting invariant curve with negative Lyapunov exponent.
In all these cases $E^s=\mathbb T\times \mathbb R$ and $E^u=\emptyset$.

The case of non-attracting invariant curves, $E^u\neq \emptyset$, is more
difficult to deal with. Numerical simulations are harder because they imply
developing algorithms suitable for computing the FHIM and their bifurcations.
Theoretical results must deal with the loss of regularity of the invariant
manifold and with stable and unstable noncontinuous directions.  Some results
appear in the literature, see \cite{HdLL06c, HdLL06verge, Figueras_Haro_2015}.

In \cite{Figueras_Haro_2015} the authors present a numerical study of some
possible bifurcation scenarios of quasiperiodic invariant curves in the
quasiperiodically driven standard map $(R_\omega, F): \mathbb T\times
\mathbb{R}^{2}\to \mathbb T\times \mathbb{R}^{2}$ given by
\begin{equation}
\label{eq: motivatingds}
\begin{cases}
\overline{\theta} & = \theta + \omega \quad (\textrm{mod }1)\\
\overline{x} & = x+\overline{y}\\
\overline{y} & = y-\frac{\partial W}{\partial x}(\theta, x) \\
\end{cases}
\end{equation} 
where $\omega\in\mathbb R-\mathbb Q$ and $ W\in C^1(\mathbb T\times
\mathbb{R},\mathbb{R})$.  For the $(\gamma, \kappa)$-parametric family of
$W(\theta,x)= \gamma x\sin(2\pi\theta)-\frac{\kappa}{(2\pi)^{2}}\cos(2\pi x)$,
they observe three types of bifurcations: \textit{smooth bifurcation},
\textit{spiky breakdown}, and \textit{folding breakdown}.  In this paper we
concentrate on the latter and a description of it goes as follows: There exists
a critical value $\gamma_c$ of the parameter $\gamma$ such that: 
\begin{itemize}
\item for all $\gamma < \gamma_c$ 
System \eqref{eq: motivatingds} has an FHIM given 
as the graph of a smooth function 
$K_\gamma:\mathbb T\rightarrow \mathbb R^2$. These 
FHIM satisfy that they are uniformly hyperbolic and $\dim E^u=\dim E^s=1$.

\item At $\gamma=\gamma_c$ 
there exists $\theta_0\in\mathbb T$ such that 
$\partial_\theta K_{\gamma_c}(\theta_0+k\omega)=\infty$ for all $k\in\mathbb Z$.

\item For $\gamma > \gamma_c$ there is a strange saddle: a
bounded measurable invariant object with positive Lyapunov exponent.
\end{itemize}

The dynamical system \eqref{eq: motivatingds} can be formulated in
terms of the formal Lagrangian 
\begin{equation}
\label{eq: lagrangian} L(\theta, \mathbf{x}) =
\sum_{k\in\mathbb{Z}}\dfrac{1}{2}(x_{k+1}-x_{k})^{2}-W(\theta+k\omega,x_{k})\,
.
\end{equation}
Fixing $ \theta_{0}\in \mathbb T $ orbits of System \eqref{eq:
motivatingds} correspond to stationary solutions of the gradient flow of $L$,
\begin{equation}
\label{eq: gradient}
\begin{aligned}
\dot{x}_{k} & = -\dfrac{\partial L}{\partial x_{k}}(\theta_{0},\mathbf{x}) \\ &
= x_{k+1}-2x_{k}+x_{k-1}+\dfrac{\partial W}{\partial
x}(\theta_{0}+k\omega,x_{k}) \, .
\end{aligned}
\end{equation} 
The Lagrangian \eqref{eq: lagrangian} is not unique, other possible formulations
are possible. For example, in the case of System \eqref{eq: motivatingds}
a possible Lagrangian could be the classic Frenkel-Kontorova model with 
quasiperiodic spring lengths.
However, all possible definitions define the same gradient flow 
\eqref{eq: gradient}.

In \cite{Figueras_Haro_2015} a more general form of Equation
\eqref{eq: gradient} is considered, namely the anti-integrable limit scenario
\begin{equation}
\label{eq: example}
\varepsilon(x_{k+1}-2x_{k}+x_{k-1})+V(\theta_{0}+k\omega,x_{k}) =
0,\quad\forall k\in\mathbb{Z}
\end{equation} 
with $\varepsilon\approx 0$.  Under the hypothesis of $V$ satisfying that for
any $\theta\in\mathbb T$ there is an $x\in\mathbb R$ such that $V(\theta, x)=0$
and $\partial_x V(\theta, x)\neq 0$ a milder version of the folding breakdown is
proven: There exists a one parametric family $V_\gamma$ and $0
< a < b < 1$ such that for all $0\leq \gamma\leq a$ the system \eqref{eq:
example} has an invariant FHIM, while for $b\leq \gamma\leq 1$ the system
has an invariant strange saddle.

In this paper we completely prove the existence of the folding bifurcation 
and generalize it to more general systems.

\paragraph{Structure of the paper} 
In Section \ref{section: formulation} we present the main results in this 
paper. These results are divided into two different cases depending on the 
dimension of the systems. In Sections \ref{section: The regular case} 
,\ref{section: The singular case} and \ref{section: rotation} we present the 
proofs of all results stated in Section \ref{section: formulation}. Finally, in 
Section \ref{section: final remarks} we formulate some final results which are 
of independent interest and give additional information relating to the main 
results.

\section{Formulation of the results}
\label{section: formulation}

Let $ h: \mathcal M\to \mathcal M $ be a homeomorphism of the compact space $\mathcal M$.
We consider systems of the form
\begin{equation}
\label{eq: generalequation} 
\varepsilon
Z(\theta_k, x_{k+1},x_{k},x_{k-1})+V(\theta_k, x_{k}) = 0, \quad \forall
k\in\mathbb{Z}
\end{equation} 
where $Z\in C^r(\mathcal M\times \mathbb{R}^{3}, \mathbb{R})$, $V\in
C^r(\mathcal M\times \mathbb{R},\mathbb{R})$, $r\geq 1$,
$\mathbf{x}=\left\{x_{k}\right\}_{k\in\mathbb Z}\in\ell^{\infty}(\mathbb{Z}) $,
$\left\{\theta_k\right\}_{k\in\mathbb Z}\in\mathcal M^{\mathbb Z}$ with $
\theta_{k}=h(\theta_{k-1})$ and $ \varepsilon\in\mathbb{R} $. If $Z(\theta, a, b,
c) = a+\tilde Z(\theta, b, c)$ and $\varepsilon\neq 0$ then System \eqref{eq:
generalequation} defines a dynamical system on $\mathcal M\times \mathbb R^2$
given by
\begin{equation}
\label{eq: dyn sys}
\left\{
\begin{array}{rcl}
\theta_{k+1} &=& h(\theta_k)\\
x_{k+1} &=& -\tilde Z(\theta_k, x_k, x_{k-1})-\frac{1}{\varepsilon}
V(\theta_k, x_k).
\end{array}
\right.
\end{equation}
With a slight abuse of notation, we will call $ V $ the \emph{potential}.

A key remark is that all the results that we present here are for
values of $\varepsilon$ small: the anti-integrable limit scenario.  In the
litterature there are several results on this direction, see
\cite{Aubry_Abramovici_1990, Chen_2007, Mackay_Meiss_1992,
Veerman_Tangerman_1991}. In all these papers they deal with the case that the
potential $V$ does not vanish at the anti-integrable limit. In this paper we
are able to provide proof of the existence of orbits even in the case that
the derivative of the potential vanishes with even order: $V(\theta, x)\approx
x^{2p+1}$, $p\in\mathbb N$.

The nature of the solutions of System \eqref{eq: generalequation} depends
heavily on $V$. Our first result and its proof are an immediate generalization
of a result appearing in \cite{Figueras_Haro_2015} and are included for
completeness. 
\begin{bigthm}
\label{thm: regular-graph} Suppose that $Z\in C^r(\mathcal M\times \mathbb R^3,
\mathbb R)$, $V\in C^r(\mathcal M\times \mathbb R, \mathbb R)$, $r\geq 1$,
satisfies that $\partial_x V(\theta, x)\neq 0$ for all $(\theta, x)$ in a
connected and bounded subset of the zero level set $V^{-1}(0)$.  Then, there
exists $\varepsilon_0 > 0$ and $ K\in C^{r}((-\varepsilon_0,
\varepsilon_0)\times \mathcal M, \mathbb R) $ such that 
\begin{equation}
\label{eq: functionalequation} 
\varepsilon Z(\theta, K(\varepsilon, h(\theta)),
K(\varepsilon, \theta), K(\varepsilon, h^{-1}(\theta)))
+V(\theta, K(\varepsilon, \theta))=0
\end{equation} 
holds for any $ \theta\in \mathcal M$ and $|\varepsilon| < \varepsilon_0$. 
Moreover, in the case that $Z$ defines a dynamical system as in \eqref{eq: 
dyn sys} then it has positive Lyapunov exponents.
\end{bigthm}

Notice that Theorem \ref{thm: regular-graph} implies that there exists bounded
orbits lying on a smooth manifold: $x_k = K(\varepsilon, h^k(\theta_0))$.

The following results show that for more general $V$'s there are still bounded
solutions. We present them for two different cases: the one dimensional and the
two dimensional cases. We decided to present them separately because, although
the statements are quite similar, the proofs are different. Moreover, this
differentiation between cases is very natural, see Section \ref{section:
introduction}. Note that the one dimensional case can be considered as a
variation of the two dimensional case satisfying $
\partial_{c}Z(\theta,a,b,c)\equiv 0 $.

\begin{rem}
	Throughout the rest of the paper the homeomorphism $ h $ is not needed and all results will remain true for general sequences $ \{\theta_{k}\}_{k\in\mathbb{Z}}\in\mathcal{M}^{\mathbb{Z}} $ with no modification of the proofs.
\end{rem}

\subsection{One dimensional case}

Let $I=[-1,1]$, $I_o=(-1,1)$, and $ V $ be a potential satisfying 
\begin{enumerate}

\labitem{(C0)}{item: Condition0} there is some $ \varepsilon_{0}>0 $ such that
for every $|\varepsilon|\le\varepsilon_{0} $ each connected component of the $
\varepsilon $-level set of $ V $ is compactly contained in $\mathcal M\times
I_{o}$ and projects surjectively onto $\mathcal M$. 
\end{enumerate}

\begin{rem}
	For the results $ \varepsilon $-level sets outside $ I $ can be allowed but restricting them avoids more cumbersome notation and makes the formulation of the results and their proofs easier.
\end{rem}

We consider here a slight variation of Equation \eqref{eq:
generalequation}. Let $Z\in C^1(\mathcal M\times\mathbb{R}^{2}, \mathbb R)$
satisfy $ Z(\mathcal M\times I^{2})\subset I_{o} $ and $ \partial_{x}Z(\theta,
x,y)\neq0 $ for every $ (\theta, x,y)\in \mathcal M\times I^{2}$. We then
consider the system
\begin{equation}
\label{eq: 1D equation} \varepsilon
Z(\theta_k,x_{k+1},x_{k})+V(\theta_k, x_{k})=0, \quad \forall k\in\mathbb{Z}
,
\end{equation} 
where $ \theta_{k} = h(\theta_{k-1})\in\mathcal M$.  Our first result establishes
the existence of solutions $ \{x_{k}\}\in
I^{\mathbb{Z}}\subset\ell^{\infty}(\mathbb{Z}) $ of Equation \eqref{eq: 1D
equation}.

\begin{bigthm}
\label{thm: existence 1d} Suppose that $ Z(\mathcal M\times I^{2})\subset I_{o}
$, $ \partial_{x}Z(\theta, x,y)\neq 0 $ for every $ (\theta, x,y)\in\mathcal
M\times I^{2}$, and $ V \in C^{1}(\mathcal M\times \mathbb R, \mathbb R) $
satisfies that there are $ -1<t_{0}<t_{1}<1 $ and an $ \varepsilon_{0}>0 $ such
that $ V^{-1}( [-\varepsilon_{0},\varepsilon_{0}])\subset \mathcal M\times
[t_{0},t_{1}]$ and projects surjectively onto $ \mathcal M $. Let $ h:\,
\mathcal{M}\to\mathcal{M} $ be a map. Then for each $
|\varepsilon|<\varepsilon_{0} $ and $\left\{\theta_k\right\}_{k\in\mathbb Z}\in
\mathcal{M}^{\mathbb{Z}}$, $\theta_k = h(\theta_{k-1})$, there is $
\left\{x_{k}\right\}_{k\in\mathbb Z}\in I^{\mathbb{Z}} $ satisfying Equation
\eqref{eq: 1D equation}.
\end{bigthm}

Define the functions $f_{\theta}(x,y)=\varepsilon Z(\theta, x,y)+ V(\theta,
y)$. The hypothesis $\partial_x Z \neq 0$ implies that $f_{\theta}^{-1}(0)$ is a one
dimensional submanifold of $ \mathbb{R}^{2} $ for every $ \theta\in \mathcal M
$. If it is compact then by the classification of one dimensional manifolds
each of its connected components must be diffeomorphic to either $ [0,1] $ or
the circle $ \mathbb T $. We give a special name to a certain class of such
submanifolds which will be of special importance for us.

\begin{defi}
A connected component of $ f_{\theta}^{-1}(0)\cap I^{2} $ is called
\textit{almost horizontal} if it is diffeomorphic to $ [0,1] $ with boundary
points $ p_{1}\in \{-1\}\times I $ and $ p_{2}\in \{1\}\times I $ and if $
p_{1},p_{2} $ are its only points of intersection with the boundary of $ I^{2}
$.
\end{defi}

\begin{defi}
Given $\mathcal M\times\mathbb R^d$, the \textit{fiber} of $\theta_0\in\mathcal
M$ is the set $\left\{\theta_0\right\}\times\mathbb R^d$.
\end{defi}

\begin{bigthm}
\label{thm: cantor 1d} Under the same assumptions as in Theorem \ref{thm:
existence 1d} and, in addition, 
assuming that for $|\varepsilon| > 0$ and $0 < \delta < 1$ there is 
$\left\{\theta_k\right\}_{k\in\mathbb Z}\in \mathcal{M}^{\mathbb{Z}}$, $\theta_k =
h(\theta_{k-1})$, satisfying the following:
\begin{enumerate}
\item The fiber over each $ \theta_{k} $ contains an almost horizontal component
with slope of absolute value at most $ 1-\delta $ everywhere,

\item Infinitely many $ \theta_{k} $'s have fibers containing at least two
almost horizontal components with slope of absolute value at most $ 1-\delta
$ everywhere.
\end{enumerate}
Then for each $ k\in\mathbb{Z} $ the coordinates $ x_{k} $ of all orbits of
Equation \eqref{eq: 1D equation} contained in almost horizontal components form
a Cantor set.
\end{bigthm}

\subsection{Two dimensional case}

In the two dimensional case we need that 
$ Z $ and $ V $ satisfy Condition \ref{item: Condition0} and the following
two:
\begin{enumerate}
\labitem{(C1)}{item: Condition1} $Z(\mathcal M\times I^{3})\subset I_{o}$,

\labitem{(C2)}{item: Condition2} 
$\frac{\partial Z}{\partial x}(\theta,x,y,z)\neq 0 \neq \frac{\partial Z}
{\partial z}(\theta, x,y,z)$  everywhere on $ \mathcal{M}\times \RR^{3} $.
\end{enumerate}
Note that Condition \ref{item: Condition1} is just a matter of scaling; since
$\mathcal M\times I^{3}$ is compact any continuous function $ f:\, \mathcal
M\times \mathbb{R}^{3}\to\mathbb{R} $ can be multiplied by a constant $ c>0 $
such that $ cf(\mathcal M\times I^{3})\subset I_{o} $. 
Note also that for $ \varepsilon
= 0 $ Equation \eqref{eq: generalequation} reduces to the equation 
\[
V(\theta_k, x_{k}) = 0, \quad \forall k\in\mathbb{Z} 
\] 
which, by Condition \ref{item: Condition0}, has a solution
$\left\{x_{k}\right\}_{k\in\mathbb Z}\in I^{\mathbb{Z}} $ for any $
\theta_{0}\in \mathcal M$.

\begin{bigthm}
\label{thm: existence 2d} 
Let $Z\in C^{1}(\mathcal{M}\times\RR^{3},\RR)$ and
$V\in C^1(\mathcal M\times\mathbb{R}, \mathbb{R})$ satisfying Conditions
\ref{item: Condition0}, \ref{item: Condition1} and \ref{item: Condition2} and
let $ h:\, \mathcal{M}\to \mathcal{M} $ be a homeomorphism.  Then for each $
|\varepsilon|<\varepsilon_{0} $ and $\theta_{0}\in \mathcal M $ there exists a
sequence $ \left\{x_{k}\right\}_{k\in\mathbb{Z}}\in I^{\mathbb{Z}} $ satisfying
Equation \eqref{eq: generalequation}.
\end{bigthm}

As in the one dimensional case we introduce the functions $
f_{\theta}(x,y,z)=\varepsilon Z(\theta,x,y,z)+V(\theta,y) $. By Condition
\ref{item: Condition2} the set $ f_{\theta}^{-1}(0) $ is a two dimensional
submanifold for each $ \theta\in\mathcal{M} $. We again give a special name to
a certain class of such submanifolds.

\begin{defi}
A connected component of $ f_{\theta}^{-1}(0)\cap I^{3} $ is called
\textit{almost horizontal} if it projects surjectively onto $ I^{2}=\{ (x,z)
\} $, it is diffeomorphic to $ I^{2} $ and if its boundary is entirely
contained inside the boundary of $ I^{3} $ and is the only intersection with
the boundary of $ I^{3} $.
\end{defi}

Furthermore each set $ f_{\theta}^{-1}(0) $ is transversal to any $ x=c $ or $
z=c $ plane for any $ c\in I $. This defines two foliations of $
f_{\theta}^{-1}(0) $ that we call the \textit{natural foliations}. The leaves
of the natural foliations correspond to almost horizontal components of one
dimensional problems.

\begin{bigthm}
\label{thm: cantor 2d} Under the same conditions as in Theorem \ref{thm:
existence 2d} and, in addition, 
assuming that for 
$|\varepsilon|>0 $ and $ 0<\delta<1 $ 
there is $\left\{\theta_k\right\}_{k\in\mathbb Z}\in\mathcal M^{\mathbb Z}$, 
$\theta_k = h(\theta_{k-1})$, satisfying the following:
\begin{enumerate}
\item The fiber over each $ \theta_{k} $ contains an almost horizontal component
such that each leaf of the natural foliations of the component has slope of
absolute value at most $ 1-\delta $ everywhere.

\item Infinitely many $ \theta_{k} $ have fibers containing at least two almost
horizontal components such that each leaf of the natural foliations of the
components has slope of absolute value at most $ 1-\delta $ everywhere.
\end{enumerate} 
Then for each $ k\in\mathbb{Z} $ the coordinates $ x_{k} $ of all orbits of
Equation \eqref{eq: generalequation} contained in almost horizontal components
form a Cantor set.
\end{bigthm}
	
\begin{rem}
The same ideas that prove the existence of orbits for the $ 2 $-dimensional
case generalize to the case $ \varepsilon
Z(\theta_k, x_{k+1},x_{k},x_{k-1},\dots,x_{k-l})+V(\theta_k, x_{k}) $ with
appropriately modified assumptions on $ Z $. The proofs for the structure of
the orbit set also go through with the straightforward generalizations though
the Cantor set disappears. Instead, we will have that the corresponding sets $
W_{+} $ and $ W_{-} $ are essentially transversal $ k-1 $-dimensional subsets
of the $ k $-dimensional submanifold $ f_{\theta_{k}}^{-1}(0) $. Thus the orbit
set would, informally, have topological dimension $ k-2 $. However, these would
have a Cantor-like distribution in $ f_{\theta_{k}}^{-1}(0) $.
\end{rem}

\subsection{Bifurcation diagram}

By Theorem \ref{thm: regular-graph} the solution set of Equation \eqref{eq:
generalequation} with a $V$ satisfying $\partial_x V\neq 0$ in 
its $0$-level set is 
a graph. On the other hand, Theorems \ref{thm: cantor 1d} and \ref{thm: cantor
2d} also show that all coordinates of certain types of solutions are contained
in Cantor sets. For families of potentials $ V_{t} $ ranging from non-degenerate to
those satisfying the conditions of Theorem \ref{thm: cantor 2d}, e.g.
potentials with a folded $ 0 $-level set, the solution set of Equation
\eqref{eq: generalequation} must undergo a bifurcation. In this section we
discuss this bifurcation. We begin by making a definition.

\begin{defi}
The potential $ V $ is called \textit{admissible} if for every $ \theta\in
\mathcal M $ there is a point $ (\theta, y)\in V^{-1}(0) $ such that $
\partial_{y}V(\theta, y)\neq 0 $.
\end{defi}

The set of admissible potentials contains the potentials with folded $ 0
$-level set, which are our prototypical example of a potential with a Cantor
set of solutions to Equation $ \eqref{eq: generalequation} $ and are the
subject of this discussion.

\begin{figure}[H]
\centering
\begin{subfigure}[b]{.3 \linewidth}
\begin{tikzpicture}[scale=.5]
\draw (-3,0) .. controls (-1,0) and (-1,-2) .. (1,0)
(1,0) .. controls (3,2) and (3,0) .. (5,0);
\draw (-3,-3) -- (-3,3) -- (5,3) -- (5,-3) -- (-3,-3);
\node at (1,-3.3) {$ \mathbb T $};
\node at (-3.3,0) {$ I $};
\end{tikzpicture}
\caption{A non-degenerate $ 0 $-level set of $ V $ for $ \mathcal M=\mathbb T $.}
\end{subfigure}
\begin{subfigure}[b]{.3\linewidth}
\begin{tikzpicture}[scale=.5]
   \draw (-3,0) .. controls (0,0) and (6,-3) .. (1,0)
      (1,0) .. controls (-4,3) and (2,0) .. (5,0);
   \draw (-3,-3) -- (-3,3) -- (5,3) -- (5,-3) -- (-3,-3);
   \node at (1,-3.3) {$ \mathbb T $};
   \node at (-3.3,0) {$ I $};
\end{tikzpicture}
\caption{An admissible folded $ 0 $-level set of $ V $ for 
$ \mathcal M=\mathbb T $.}
\end{subfigure}
\begin{subfigure}[b]{.3\linewidth}
\begin{tikzpicture}[scale=.5]
\draw[smooth] (-3,0) .. controls (0,0) and (1,-4) .. (1,0)
(1,0) .. controls (1,4) and (2,0) .. (5,0);
\draw (-3,-3) -- (-3,3) -- (5,3) -- (5,-3) -- (-3,-3);
\node at (1,-3.3) {$ \mathbb T $};
\node at (-3.3,0) {$ I $};
\end{tikzpicture}
\caption{A nonadmissible $ 0 $-level set of $ V $ for $\mathcal M=\mathbb T $.}
\label{fig: nonadmissible}
\end{subfigure}
\caption{}
\end{figure}

Note that any family of potentials going from non-degenerate to folded must pass
through a nonadmissible potential as shown in Figure \ref{fig: nonadmissible}.
The bifurcation happens around this transition. The following explicit example
of such a family is given for $ \mathcal M=\mathbb T $ in
\cite{Figueras_Haro_2015}:

\[ V_{s}(\theta,x)=(x^{2}+a(\theta))(x-b(\theta))+2.15-0.15s \] 
where $
a(\theta)=1.1-1.2\sin(2\pi(\theta+0.2)) $ and $
b(\theta)=1.2+1.2\cos^{2}(\pi\theta) $. For $ s=0 $ the potential is admissible
with a folded $ 0 $-level set and for $ s=1 $ the potential is non-degenerate. As such
the solution set to $ \varepsilon Z+V_{s}=0 $ undergoes such a bifurcation as $
s $ goes from $ 1 $ to $ 0 $ for any $ Z $ satisfying the general conditions
and any $ \varepsilon $ small enough.

\begin{rem}
In the case that $Z$ defines a dynamical system we can talk about the stability
of the orbits proven in the previous theorems.  Following the results on
stability in \cite{Figueras_Haro_2015} we obtain that in the two-dimensional
case the solutions on both the smooth manifold, Theorem \ref{thm:
regular-graph}, and on the Cantor set, Theorem \ref{thm: cantor 2d}, have
positive Lyapunov exponent and one dimensional stable and unstable bundles. In
the one-dimensional case, both theorems \ref{thm: regular-graph} and \ref{thm:
cantor 1d} imply that the solutions have positive Lyapunov exponent: they are
repellers.
\end{rem}

\subsection{Fibered rotation numbers and commensurate 
and incommensurate Cantor Sets}

The forward fibered rotation number of a sequence
$\left\{x_k\right\}_{k\in\mathbb Z}$ is, if it exists, defined by 
\[
\rho=\lim_{N\rightarrow +\infty}\frac1N \sum_{k=0}^{N-1} x_{k+1}-x_k.
\]
The backward fibered rotation number is defined analogously by computing the
limit for $N\rightarrow -\infty$. If both the forward and backward coincide,
then its called the fibered rotation number of the sequence.

By a slightly change on the proof of Theorems \ref{thm: existence 1d} or
\ref{thm: existence 2d} we can prove the existence of orbits with any desired
rotation number. 
\begin{bigcor}
\label{cor: AM set} Under the assumptions of Theorem \ref{thm: existence 1d} or
\ref{thm: existence 2d} and, if $Z$ and $V$ satisfy $Z(\theta, a, b,
c)=Z(\theta, a+1, b+1, c+1)$ and $V(\theta, x)=V(\theta, x+1)$ for every $a, b,
c, x\in\mathbb R$, and $Z(\mathcal M\times [-2,2]^3)\subset [-1,1]$, then for
every $\omega \in\mathbb R$ System \eqref{eq: generalequation} admits a
solution $\left\{y_k\right\}_{k\in\mathbb Z}$ with fibered rotation number
$\omega$ and satisfying 
\begin{equation}
\label{eq: inequality}
|y_k-k\omega|\leq 2.
\end{equation}
\end{bigcor}

The result in Corollary \ref{cor: AM set} goes in the same lines 
as the ones appearing in \cite{Mackay_Meiss_1992} but with a very 
remarkable difference: our results also apply for standard symplectic 
maps with potentials having vanishing derivatives. This could lead 
to the existence of this kind of sets with zero Lyapunov exponents, but 
we have not explored this possibility in this paper. 

It is also worth noticing that Corollary \ref{cor: AM set} generalizes 
the results on the existence of Aubry-Mather sets appearing in 
\cite{delaLlave_Valdinoci_2007}. It could be of interest to explore if 
with the variational techniques developed there similar results as 
Corollary \ref{cor: AM set} could be derived.

\section{The non-degenerate case}
\label{section: The regular case}

\begin{proof}[Proof of Theorem \ref{thm: regular-graph}]
Consider the smooth functional $ \mathcal{F}:C^r(\mathcal M, \mathbb R)\times
\mathbb{R}\to C^r(\mathcal M, \mathbb R)$ given by
\begin{equation}
\mathcal{F}(f,\varepsilon)(\theta) = \varepsilon
Z(\theta, f(h(\theta)),f(\theta),f(h^{-1}(\theta)))+V(\theta, f(\theta)) \, .
\end{equation} 
By hypothesis there is a function $ f_{0}\in C^r(\mathcal M, \mathbb R)$
satisfying $ \mathcal{F}(f_{0},0)\equiv 0 $ with  $D_1\mathcal F(f_0, 0)$
invertible. Hence, by the Implicit Function Theorem there are neighborhoods $
U_{1} $ of $ f_{0} $ in $ C^{r}(\mathcal{M}, \mathbb R) $, $ U_{2} $ of $ 0 $
in $ \mathbb{R} $ and a continuous map $ \mathbb{R}\to C^{r}(\mathcal{M},
\mathbb R) $ sending $ \varepsilon\mapsto f_{\varepsilon} $ such that $
\mathcal{F}(f,\varepsilon)\equiv 0 $ if and only if $ f=f_{\varepsilon} $ in
these neighborhoods. Thus we can set $
K(\varepsilon,\theta)=f_{\varepsilon}(\theta) $.
\end{proof}

\section{The degenerate case}
\label{section: The singular case}

We now consider the more general case where we do not put any conditions on $
\partial_{x}V $.

\subsection{One dimensional case}

\begin{proof}[Proof of Theorem \ref{thm: existence 1d}]
The proof of the existence of the solutions $ \left\{x_{k}\right\}_{k\in\mathbb
Z} $ is essentially based on controlling the preimages of the associated 
dynamical system.

Let $ k\in\mathbb{Z} $ and let some $ x_{k+1}\in I $ be given. 
By the Intermediate Value Theorem equation
\[ 
\varepsilon Z(\theta_{k},x_{k+1},x)+V(\theta_{k},x)=0 \,  
\] 
has a solution $x_k\in I$. Recursively, the equations
\[ 
\varepsilon Z(\theta_{k-l},x_{k-l+1},x)+V(\theta_{k-l},x)=0
\,  \]
have solutions $x_{k-l}\in I$ for any $l > 0$.

To proceed, define the sets 
\[ 
B_{k} = \{ \left\{x_{m}\right\}_{m\in\mathbb Z}\in I^{\mathbb{Z}}:\,
\varepsilon Z(\theta_{k},x_{k+1},x_{k})+V(\theta_{k},x_{k})=0 \} \, . 
\] 
Note that $ B_{k} $ is a closed subset of $ I^{\mathbb{Z}} $ for each $ k $
since $ \varepsilon Z+V $ is a continuous function and hence its $ 0 $ level
set is closed. The sequence of sets $ B_{k} $ satisfy the finite intersection
property. To see this, note first that $ B_{k}\cap B_{l}\neq\emptyset $ follows
immediately for $ |k-l|\ge 2 $. To see that $ B_{k}\cap B_{k-1}\neq \emptyset $
we can use the idea from above that given $ x_{k+1},x_{k}\in I $ satisfying $
\varepsilon Z(\theta_{k},x_{k+1},x_{k})+V(\theta_{k},x_{k})=0 $ we can always
find $ x_{k-1}\in I $ such that $ \varepsilon
Z(\theta_{k-1},x_{k},x_{k-1})+V(\theta_{k-1},x_{k-1})=0 $. From this we can get
a sequence $\left\{x_{m}\right\}_{m\in\mathbb Z}\in B_{k}\cap B_{k-1} $ by
picking any $ x_{l}\in I $ for $ l\neq k+1, k, k-1 $. By induction it follows
that any intersection of the form
\[ 
\bigcap_{k_{1}\le k\le k_{2}}B_{k} 
\] 
is also nonempty for $ k_{1}\le k_{2}
$. Since the intersection of any finite subcollection must contain such an
intersection the finite intersection property follows.

Now since $ I^{\mathbb{Z}} $ is compact by Tychonoff's theorem we get that the
intersection \[ B=\bigcap_{k\in\mathbb{Z}}B_{k} \] is nonempty and hence
contains the solutions $\left\{x_{k}\right\}_{k\in\mathbb
Z}\in I^{\mathbb{Z}} $ that we are looking for.
\end{proof}

\begin{proof}[Proof of Theorem \ref{thm: cantor 1d}]
Begin by fixing any $ k\in\mathbb{Z} $. Let $ (x_{k+1},x_{k})\in I^{2} $ be a
solution of 
\[
f_{\theta_{k}}(x_{k+1}, x_k)=\varepsilon Z(x_{k+1}, x_k)+V(\theta_k, x_k) =0
\]
contained inside an almost horizontal component. Since $
\partial_{x_{k+1}}f_{\theta_{k}}\neq 0 $ we can use the implicit function
theorem to find a surjective function $ x_{k+1}(x_{k}) $ such that $
f_{\theta_{k}}(x_{k+1}(x_{k}),x_{k})=0 $ for every $ x_{k} $ in some closed set
$ O_{1}\subset I$. Similarly, for $ f_{\theta_{k+1}}^{-1}(0) $ we can also find
an almost horizontal component and corresponding surjective function $
x_{k+2}(x_{k+1}):\, \widetilde{O}_{2}\to I $. By composition we thus get a
surjective function $ x_{k+2}(x_{k}):\, O_{2}\to I $ where $ O_{2}\subset O_{1}
$. Continuing in this fashion we get a sequence of closed sets $ \dots\subset
O_{n}\subset O_{n-1}\subset\dots O_{2}\subset O_{1}\subset I $ and
corresponding functions $ x_{k+n}(x_{k}):\, O_{n}\to I $.

Now consider some fixed set $ O_{n} $. If the fiber over $ \theta_{n+1} $ has
more than one almost horizontal component then there would be more than one
choice of $ O_{n+1} $, call them $ O_{n+1,j_{n+1}} $ where $ j_{n+1} $ are
indexed by a finite set $ J_{n+1}\subset \mathbb N $. Thus $ O_{n} $ can be
subdivided into $ |J_{n+1}| $ connected components. Similarly, if the fiber
over $ \theta_{n+1} $ has only one almost horizontal component then $
|J_{n+1}|=1 $. 

Since by assumption there are infinitely many $ n\in\mathbb{Z} $ such that $
|J_{n}|\ge 2 $ this allows for the construction of a Cantor set. Note that at
level $ n $ the total number of connected components of points solving $
f_{\theta_{k}}=0,\dots,f_{\theta_{k+n}}=0 $ is $ \prod_{i=1}^{n}|J_{i}|<\infty
$. Denote the complete level $ n $ set by $ \overline{O}_{n} $. Then $
\overline{O}_{n} $ is closed since it is a finite union of closed sets. It is
compact since it is a subset of $ I $ and the sequence of sets $
\left\{\overline{O}_{n}\right\}_{n\in\mathbb N} $ is nested such that $
\overline{O}_{n+1}\subset\overline{O}_{n} $. By the finite intersection
property it is therefore nonempty. Let $ W=\bigcap_{n\ge 1}\overline{O}_{n} $.
Then $ W $ is metrizable since it is a subset of a metric space. In order to
show that it is a Cantor set it therefore remains to show that it has no
isolated points and that it is totally disconnected.

Let $ x\in W $ and $ N $ be an open neighborhood of $ x $. At each level
of the construction $ x $ belongs to some pulled back almost horizontal
component. From the bound on the slope of the almost horizontal components
we get that for any $ n $ \[ |O_{n}|\le (1-\delta)|O_{n-1}| \] and by
iteration we get that $ |O_{n}|\le 2(1-\delta)^{n} $. Hence we have that for
sufficiently large $ n $ there must be an entire component of the $
\overline{O}_{n} $ contained inside $ N $. Furthermore, since there are
infinitely many $ i $ for which $ \theta_{i} $ has at least $ 2 $ surjective
components this component contained inside $ N $ must eventually split into
at least $ 2 $ components. Each of these components must contain points of $
W $ and therefore $ W $ cannot be connected and $ x $ cannot be isolated.
\end{proof}

\begin{rem}
   The construction of the sets $ O_{n} $ from the first paragraph of the above
   proof can also be used to show existence of solutions.
\end{rem}

Before leaving the one dimensional case for the two dimensional case we prove
the following one dimensional lemma that is useful for the two dimensional
case.

\begin{lem}
\label{lem: almosthorizontalcomponent} Let $ V\in C^{r}(\mathcal{M}\times\mathbb{R},\mathbb{R}) $ satisfy Condition \ref{item: Condition0} and let $ Z\in C^{r}(\mathcal{M}\times\mathbb{R}^{2},\mathbb{R}) $, $ r\ge 1 $, satisfy 
\begin{itemize}
	\item $ Z(\mathcal{M}\times I^{2})\subset I_{o} $,
	\item $ \partial_{x}Z(\theta,x,y)\neq 0 $ for every $ (\theta,x,y)\in\mathcal{M}\times\mathbb{R}^{2} $.
\end{itemize}
Then for every $ \theta\in\mathcal{M} $ and every $ 0<|\varepsilon|<\varepsilon_{0} $ the set $ f_{\theta}^{-1}(0)\cap I^{2} $ contains an almost horizontal component.
\end{lem}

\begin{figure}[H]
	\centering
	\begin{tikzpicture}
		\draw[scale=1,domain=-0.9:0.9, smooth, variable=\x] plot ({1/(\x*\x-1)+1.5},{\x+1}) node[xshift=1.5 cm, yshift=.3 cm] {$ f_{\theta}=0 $};
		\draw[scale=1,domain=-0.9:0.9, smooth, variable=\x] plot ({-1/(\x*\x-1)-1.5},{\x-1}) node[xshift=-1.5 cm, yshift=-2.2 cm] {$ f_{\theta}=0 $};
		\node at (-2,1) {$ f_{\theta}<0 $};
		\node at (2,1) {$ f_{\theta}>0 $};
		\node at (-2,-1) {$ f_{\theta}<0 $};
		\node at (2,-1) {$ f_{\theta}>0 $};
		\draw[dashed] (-3.76,-3.76) rectangle (3.76,3.76);
		\node at (4.3,1) {$ I^{2} $};
	\end{tikzpicture}
	\caption{Impossibility of two components of $ f_{\theta}^{-1}(0) $ intersecting $ (\partial I)\times I_{0} $ on opposite sides with no almost horizontal component for the case $ \partial_{x}Z>0 $. For $ \partial_{x}Z<0 $ the signs are reversed.}
	\label{fig: asymptoteimpossible}
\end{figure}
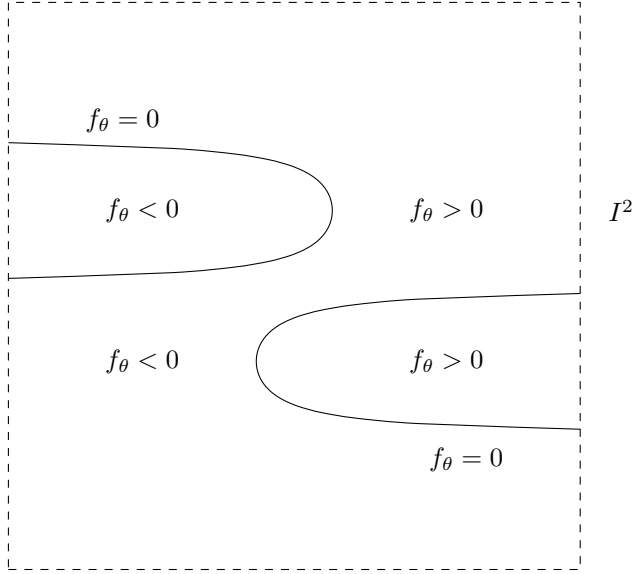

\begin{proof}
We begin by noting that $ 0 $ is a regular value of $ f_{\theta} $ and so $
f_{\theta}^{-1}(0) $ is a smooth one dimensional manifold and $
f_{\theta}^{-1}(0)\cap I^{2} $ is compact. Furthermore, since $
(x,y)\in f_{\theta}^{-1}(0)\cap I^{2} $ implies $ (\theta,y)\in
V^{-1}([-\varepsilon_{0},\varepsilon_{0}])\subset \mathcal{M}\times
[t_{0},t_{1}] $ we also have that each component of $ f_{\theta}^{-1}(0)\cap
I^{2} $ is in fact a smooth, compact submanifold of $ I\times I_{o} $
contained in $ I\times[t_{0},t_{1}] $. By the classification of one
dimensional, smooth, compact manifolds we then have that $
f_{\theta}^{-1}(0)\cap I^{2} $ is diffeomorphic to a finite union of 
circles, line segments and isolated points. It must also project surjectively onto the $ x
$-axis since for any $ x $ Equation \eqref{eq: 1D equation} can be solved
for $ y $ by the intermediate value theorem. It remains to prove 
that the only possible case is the almost horizontal one.
	
First we note that no connected component of $ f_{\theta}^{-1}(0) $ can be
diffeomorphic to a circle since it cannot have any horizontal tangencies by the
condition $ \partial_{x}Z(\theta,x,y)\neq 0 $.
	
Next we note that any isolated points or the endpoints of any component diffeomorphic
to a line segment must be contained in $ (\partial I)\times I_{0} $ by the fact
that it must be contained inside $ I\times[t_{0},t_{1}] $ and that $
f_{\theta}^{-1}(0) $ is a smooth one dimensional submanifold of $
\mathbb{R}^{2} $ without boundary.
	
It also follows that $ f_{\theta}^{-1}(0)\cap I^{2} $ must have at least one
component which is diffeomorphic to a line segment. If this component
intersects both the left boundary $ \{-1\}\times I_{o} $ and the right boundary
$ \{1\}\times I_{o} $ we have an almost horizontal component and we are done.
Otherwise, if both endpoints are contained in one side of the boundary, there
must be another component diffeomorphic to a line segment which is either
almost horizontal or whose endpoints are contained in the other side of the
boundary in order for the set to project surjectively. In the first case we are
again done. In the second case each of the curves divide the square $ I^{2} $
into two parts: the inside, whose boundary is formed by the curve itself and
the line segment connecting its endpoints, and the complementary outside. The
values of $ f_{\theta} $ on the inside and outside of such a curve differ by
sign. This leads to a contradiction as on the part of $ I^{2} $ outside both
curves $ f_{\theta} $ would have to take on values of both signs but never
zero, see Figure \ref{fig: asymptoteimpossible} for an illustration in the case
$ \partial_{x}Z>0 $. Thus there must be an almost horizontal component even in
this case.  
\end{proof}

\begin{rem}
In fact it can be shown that every connected component of the intersection of $
V^{-1}([-\varepsilon_{0},\varepsilon_{0}]) $ with the fiber over $ \theta\in M
$ on which $ V(\bullet,\theta) $ is surjective onto $
[-\varepsilon_{0},\varepsilon_{0}] $ must contain an almost horizontal
component. 
\end{rem}

\subsection{Two dimensional case}

We suppose that $ Z $ and $ V $ are functions satisfying Conditions \ref{item:
Condition0}, \ref{item: Condition1} and \ref{item: Condition2}, from the
introduction but will only consider $ \varepsilon\neq 0 $. We will prove the
following.

\begin{proof}[Proof of Theorem \ref{thm: existence 2d}]
For a fixed $\theta_{0}\in \mathcal{M}$ and $l> 0$ consider the finite
dimensional system of equations
\[
f_{k}(x_{k+1},x_{k},x_{k-1}) = \varepsilon
Z(\theta_{k},x_{k+1},x_{k},x_{k-1})+V(\theta_{k},x_{k})
\]
for $-l\leq k \leq l$, with $x_{l+1}=a\in I$ and $x_{-l-1}=b\in
I$. Since 
\begin{equation}
\begin{aligned}
\partial_{x_{l-1}}f_{l}(a,x_{l},x_{l-1})\neq 0 \, ,
\end{aligned}
\end{equation} 
for any $ x_{l-1}\in I $ we can find $
x_{l}\in I $ such that $ f_{l}(a,x_{l},x_{l-1})=0 $ by our general
assumptions. By the implicit function theorem we can thus find open sets $
U_{l}\subset I $ and $ O_{l}\subset I $ along with a $ C^{1} $ function $
x_{l-1}(x_{l}):U_{l}\to O_{l} $ such that 
\begin{equation}
f_{l}(a,x_{l},x_{l-1}(x_{l}))=0
\end{equation} 
for any $ x_{l}\in U_{l} $. Note that $ f_{l}^{-1}(0)\cap I^{2} $ is a
submanifold of $ I_{o}\times I $, the first component corresponding to
$ x_{l} $ and the second one to $ x_{l-1} $. $ f^{-1}_{l}(0) $ projects
surjectively onto the $ x_{l-1} $ axis. In this setting we can apply 
Lemma \ref{lem: almosthorizontalcomponent} to show the
existence of an \textit{almost horizontal curve} in $ \{ (x_{l},x_{l-1})\in
I^{2} : f_{l}(a,x_{l},x_{l-1})=0 \} $ and we can therefore take $ O_{l}=I $
and $ x_{l-1}(x_{l}): U_{l}\to I $ surjective. We use the word curve here to
distinguish it from an almost horizontal component in the two dimensional case.

Next consider the equation $ f_{l-1}(x_{l},x_{l-1}(x_{l}),x_{l-2})=0 $ with
$ x_{l} $ restricted to $ U_{l} $. By the intermediate value theorem we then 
have that for any $x_{l-2} $ there is an $ x_{l} $ such that \[
f_{l-1}(x_{l},x_{l-1}(x_{l}),x_{l-2})=0 \, . \] Using the implicit function
theorem and Lemma \ref{lem: almosthorizontalcomponent} again we can find a 
closed, connected set $ U_{l-1}\subset U_{l} $ and a surjective $
C^{1} $ function $ x_{l-2}(x_{l}):U_{l-1}\to I $.

Now consider $ f_{l-2} $ with $ x_{l-1}(x_{l}) $ and $ x_{l-2}(x_{l}) $, and $
x_{l}\in U_{l-1} $. Then for any $ x_{l-3} $ we can find, just as above, an $
x_{l}\in U_{l-1} $ such that \[ f_{l-2}(x_{l-1}(x_{l}),x_{l-2}(x_{l}),x_{l-3})
= 0 \, . \] Thus we also find a closed, connected set $ U_{l-2}\subset U_{l-1}
$ and a $ C^{1} $ surjection $ x_{l-3}(x_{l}):U_{l-2}\to I $.

Proceeding by induction we can find a nested sequence of closed set $
U_{-l+1}\subset U_{-l+2}\subset \dots\subset U_{l-1}\subset U_{l} $ and
corresponding $ C^{1} $ surjections $ x_{k}(x_{l}):U_{k+1}\to I $. Lastly,
consider $ f_{-l}(x_{-l+1},x_{-l},b) $. By one final application of the
intermediate value theorem we find an $ x_{l} $ such that \[
f_{-l}(x_{-l+1}(x_{l}),x_{-l}(x_{l}),b)=0 \, . \] We have thus created an orbit
of $ f_{k}=0 $ for all $ -l\le k\le l $ for any boundary conditions $
x_{l+1}=a,\, x_{-l-1}=b $. Denote the set of all such orbits by $ S^{a,b}_{l} $
and let $ S_{l}=\bigcup_{a,b\in I}S^{a,b}_{l} $. As done in Theorem \ref{thm:
cantor 1d}, this can be considered a closed subset of the compact space $
I^{\mathbb{Z}} $. We then have $ S_{l+1}\subset S_{l} $ so the sequence of
closed sets $ \left\{S_{l}\right\}_{l\in\mathbb N}$ is nested and, by above,
each $ S_{l} $ is nonempty.  Thus the sequence of sets has the finite
intersection property and therefore the set $
S_{\infty}=\bigcap_{l=0}^{\infty}S_{l} $ is nonempty and contains full
orbits $\left\{x_{k}\right\}_{k\in\mathbb Z}\in I^{\mathbb{Z}} $ satisfying
Equation \eqref{eq: generalequation}.
\end{proof}

\begin{rem}
Though the above proof only applies to the case $ \varepsilon\neq 0 $
it is easy to find solutions for $ \varepsilon=0 $ as well since the $ x_{k}
$'s decouple, i.e. they are independent of each other, and by the general
assumptions $ V(\theta,\bullet) $ has at least one zero for every $
\theta\in \mathcal{M} $. Furthermore these are contained in $ I $.
\end{rem}

\begin{rem}
There could be orbits which also pass through components which only
intersect one boundary component of $ I^{2} $. We will not examine these in
this paper.
\end{rem}
						
\begin{proof}[Proof of Theorem \ref{thm: cantor 2d}]
Fix $ k\in\mathbb{Z} $. Using the implicit function
theorem we can, as in the proof of Theorem \ref{thm: existence 2d}, 
find a surjective function $x_{k-1}(x_{k+1},x_{k}):\, O_{-1}\to I $ 
corresponding to the almost horizontal component, where 
$ O_{-1}\subset\{(x_{k},x_{k+1})\in I^{2}\} $ has surjective projection onto 
the $ x_{k+1} $-axis, satisfying 
\[f_{\theta_{k}}(x_{k+1},x_{k},x_{k-1}(x_{k+1},x_{k}))=0 \] for every $
(x_{k+1},x_{k})\in O_{-1} $. Similarly, for $ f^{-1}_{\theta_{k-1}}(0) $ we
can find a corresponding surjective function $ x_{k-2}(x_{k},x_{k-1}):\,
\widetilde{O}_{-2}\to I $. Then, by considering the surjective map $
g_{-1}:\, O_{-1}\to I^{2} $ given by 
\[
g_{-1}(x_{k+1},x_{k})=(x_{k},x_{k-1}(x_{k+1},x_{k}))
\]
we can consider the pullback $ O_{-2}:=g_{-1}^{-1}(\widetilde{O}_{-2}) $.
Continuing inductively we can construct surjective maps $
x_{k-n}(x_{k-n+2},x_{k-n+1}):\, \widetilde{O}_{-n}\to I  $ and $ g_{-n}:\,
\widetilde{O}_{-n}\to I^{2} $. By pulling back each $ \widetilde{O}_{-n} $
through each $ g_{-n} $ in order we get sets $ O_{-n}\subset
O_{-n+1}\subset\dots\subset O_{-1} $. Note that each set in this sequence has
surjective projection onto the $ x_{k+1} $-axis.

In the same way we can also find a surjective function $
x_{k+1}(x_{k},x_{k-1}):\, O_{1}\to I $ satisfying \[
f_{\theta_{k}}(x_{k+1}(x_{k},x_{k-1}),x_{k},x_{k-1})=0 \] for every $
(x_{k},x_{k-1})\in O_{1} $. We then consider the map $ g_{1}:\, O_{1}\to
I^{2} $ given by 
\[
g_{1}(x_{k},x_{k-1})= (x_{k+1}(x_{k},x_{k-1}),x_{k}) .
\]
Thus we can consider the pullback $ O_{2}:=g_{1}^{-1}(\widetilde{O}_{2}) $.
Proceeding inductively we again get a nested sequence of sets $ O_{n}\subset
O_{n-1}\subset\dots\subset O_{1} $, each set having surjective projection
onto the $ x_{k-1} $-axis.

Now consider some set $ O_{n} $, $ |n|\ge 1 $. If the fiber over $
\theta_{k+n\pm 1} $ has more than one surjective component then there would be
more than one choice of $ O_{n\pm 1} $, call them $ O_{n\pm 1,j_{n\pm 1}} $
for $ j_{n\pm 1}\in J_{n\pm 1} $ where $ J_{n\pm 1} $ is a finite set. Thus
$ O_{n} $ can be divided into $ |J_{n\pm 1}| $ connected components. As in
the $ 1 $-dimensional case we will denote the complete level $ n $ set by $
\overline{O}_{n} $. Letting $ O_{0}=I $ and $ |J_{0}|=1 $ we can then write
the number of connected components of $ \overline{O}_{n} $ as $
\prod_{i=0}^{n}|J_{i}|<\infty $, each component projecting surjectively onto
$ x_{k+1} $ if $ n\le -1 $ and onto $ x_{k-1} $ if $ n\ge 1 $. They are also
closed and nested. 

For $ n\ge 0 $ we can thus consider the sets $
\overline{W}^{+}_{n}=\bigcap_{0\le i\le n}\overline{O}_{i} $ and $
\overline{W}^{-}_{n}=\bigcap_{-n\le i\le 0}\overline{O}_{-i} $ and their
embeddings $ W^{+}_{n},\, W^{-}_{n} $ into the almost horizontal component
given by $ W^{+}_{n}=\{ (x_{k+1}(x_{k},x_{k-1}),x_{k},x_{k}):\,
(x_{k},x_{k-1})\in \overline{W}^{+}_{n} \} $ and $ W^{-}_{n}=\{
(x_{k+1},x_{k},x_{k-1}(x_{k+1},x_{k})): (x_{k+1},x_{k})\in \overline{W}^{-}_{n}
\} $. Then each component of $ W^{+}_{n} $ projects surjectively onto the $
x_{k-1} $-axis while each component of $ W^{-}_{n} $ projects surjectively onto
the $ x_{k+1} $-axis. Since they are both contained inside a surface we must
therefore have that each connected component of $ W^{+}_{n} $ intersects every
connected component of $ W^{-}_{n} $ and vice versa. Thus we define $ W_{n} =
W^{+}_{n}\cap W^{-}_{n} $. Note that the sequence of sets $ W^{+}_{n} $ and the
sequence $ W^{-}_{n} $ are both nested and hence so is the sequence $ W_{n} $.
It follows by compactness that the set $ W=\cap_{n\ge 0}W_{n} $ is nonempty.
This is our prospective Cantor set. Note that since $ W $ is a subset of a 
metric space it is automatically
metrizable so we only have to show that it has no isolated points and that it
is totally disconnected.  To this end, let $ x=(x_{k+1},x_{k},x_{k-1})\in W $
and $ N $ be a neighbourhood of $ x $ in $ \mathbb{R}^{3} $. Pick a sequence of
sets $ O_{n} $ satisfying $ x\in O_{n} $ for every $ n\in\mathbb{Z} $. Then the
sets $ O_{n} $, as $ n\to \pm\infty $, get contracted in the $ x_{k}
$-direction by a factor $ 1-\delta $ at each level by the condition on the
natural foliations. Therefore we can find a $ n\ge 0 $ large enough so that the
$ x_{k} $ and $ x_{k-1} $ coordinates of the image of $ O_{-n} $  are contained
inside $ N $ and such that the $ x_{k} $ and $ x_{k+1} $ coordinates of the
image of $ O_{n} $ are also contained inside $ N $. Thus the connected
component of $ W_{n} $ corresponding to $ O_{n}\cap O_{-n} $ is entirely
contained inside $ N $. Furthermore both $ O_{n} $ and $ O_{-n} $ must split
into two or more connected components by assumption.  Therefore the set $ N $
must contain points of $ W $ other than $ y $ and those points must be in a
different connected component. Thus $ W $ can have no isolated points and is
totally disconnected.
\end{proof}

\section{Existence of orbits with any fibered rotation number.}
\label{section: rotation}

\begin{proof}[Proof of Corollary \ref{cor: AM set}]

We provide here the proof under the assumptions of Theorem \ref{thm: existence
2d}. The other case is done mutatis mutandis.

Fix $\omega\in\mathbb R$ and let $m_k = \lfloor k\omega \rfloor$, where
$\lfloor \alpha\rfloor=\max_{k\in\mathbb Z}\left\{k\le \alpha\right\}$. Note
that $m_{k+1} = m_k+\delta_k$ with $\delta_k\in \left\{0, 1, -1\right\}$.  For
any $a, b, c\in I$ the function $Z$ satisfies that $Z(\theta, a+m_{k-1}, b+m_k,
c+m_{k+1}) = Z(\theta, a, b, c)+ G(\theta, m_{k-1}, m_k, m_{k+1})$ with
$G:\mathcal M\times \mathbb Z^3\rightarrow \mathbb R$ satisfying
$|G|\leq 2$.
Hence we obtain the equivalent System 
\begin{equation*}
\varepsilon
\hat Z(\theta_k, x_{k+1},x_{k},x_{k-1})+V(\theta_k, x_{k}) = 0, \quad \forall
k\in\mathbb{Z}
\end{equation*}
with $\hat Z(\theta_k, a, b, c) = Z(\theta_k, a, b, c)+ G(\theta_k, m_{k-1},
m_k, m_{k+1})$. By Theorem \ref{thm: existence 2d} it has a solution
$\left\{x_k\right\}_{k\in\mathbb Z}\in I^{\mathbb Z}$.  Finally, the sequence
$\left\{y_k\right\}_{k\in\mathbb Z}$ defined by $y_k = x_k+m_k$ has fibered
rotation number $\omega$ and satisfies Inequality \eqref{eq: inequality}
because of $|m_k-m_{k-1}|\leq 1$ for all $k\in\mathbb Z$.
\end{proof}

\section{Final remarks and further related results}
\label{section: final remarks}

We present here a pair of results which are of interest both independently and in relation to the main results. 

Since Theorems \ref{thm: cantor 1d} and \ref{thm: cantor 2d} are formulated in terms of certain almost horizontal components it is of interest to know a priori if a certain system contain such components. These results give conditions on $ V $ that guarantee the existence of such almost horizontal components. In particular they show that admissible potentials satisfy the hypotheses of Theorems \ref{thm: cantor 1d} and \ref{thm: cantor 2d}.

\begin{prop}
\label{lem: 1-D surj-vert} Let $ (\theta_{0},y_{0})\in V^{-1}(0) $ such that $
\partial_{y}V(\theta_{0},y_{0})\neq 0 $. Then for every small enough $
\varepsilon $ there is a neighborhood of $ y_{0} $ in $ I $ containing an
almost horizontal component of $ f^{-1}_{\theta_{0}}(0) $ which is also a graph
over the $ x $-axis. Furthermore, for each $ 0<|\varepsilon|<\varepsilon_{0} $
the size of the projection onto the $ y $-axis of each connected component is
bounded from below by a positive constant.
\end{prop}

\begin{proof}
Consider the map $ \mathcal{F}:\,C^{1}(I,\mathbb{R})\times\mathbb{R}\to C^{1}(I,\mathbb{R}) $ given
by \[ \mathcal{F}(y,\varepsilon)(x)=\varepsilon
Z(\theta_{0},x,y(x))+V(\theta_{0},y(x)) \, . \] Then this map is Fr\'echet
differentiable. Letting $ y_{*} $ denote the constant function $ y_{*}(x)\equiv
y_{0} $ we have $ \mathcal{F}(y_{*},0)=0 $. Since $
\partial_{y}V(\theta_{0},y_{*}(x))\neq 0 $ we can apply the implicit function
theorem giving us a family of functions $ y_{\varepsilon}\in C^{1}(I) $ defined
for sufficiently small $ |\varepsilon|>0 $ such that $ \varepsilon
Z(\theta_{0},x,y(x))+V(\theta_{0},y(x))\equiv 0 $. This proves the first part
of the lemma.

For the second part we fix a small enough $ \varepsilon $ and consider the
corresponding $ y_{\varepsilon} $. By compactness we then have $
|\partial_{x}Z(\theta_{0},x,y)|\ge K_{1} $ for some constant $ K_{1}>0 $ and $
|\partial_{y}Z(\theta_{0},x,y)+\frac{1}{\varepsilon}
\partial_{y}V(\theta_{0},y)|\le K_{2} $ for some constant $ K_{2}>0 $ for every
$ (x,y)\in I^{2} $. Now pick some $ (x_{0},y_{0}) $ in an almost horizontal
component. By the implicit function theorem we can then write $ x(y) $ as a
surjective function on some neighborhood of $ y_{0} $. From the above bounds we
get that $ |x^{\prime}(y)|\le \frac{K_{2}}{K_{1}} $ and hence the neighborhood
around $ y_{0} $ must have size at least $ 2\frac{K_{1}}{K_{2}} $.  
\end{proof}
		
\begin{prop}
\label{lemma: surj-vert} Let $ (\theta_{0},y_{0})\in V^{-1}(0) $ such that $
\partial_{y}V(\theta_{0},y_{0})\neq 0 $. Then for every small enough $
\varepsilon $ there is a neighborhood of $ y_{0} $ in $ I $ containing an
almost horizontal component of $ f_{\theta_{0}}^{-1}(0) $ which is also a graph
over the $ x $-$ z $ plane. Furthermore, for each $
0<|\varepsilon|<\varepsilon_{0} $ the size of the projection onto the $ y
$-axis of each almost horizontal component is bounded from below by a positive
constant.
\end{prop}
			
\begin{proof}
Follow the proof of Lemma \ref{lem: 1-D surj-vert} with $ y=y(x,z) $ and $
\mathcal{F}:\, C^{1}(I^{2},\mathbb{R})\times\mathbb{R}\to
C^{1}(I^{2},\mathbb{R}) $. 

For the second part, use the implicit function theorem to write $ x=x(y,z) $ or
$ z=z(x,y) $. The bounds on the derivatives apply as before.
\end{proof}

\section*{Acknowledgments}
The authors want to thank professor de la Llave for his fruitful comments 
on the paper.
\bibliography{./bibliography}{}
\bibliographystyle{abbrv}
\end{document}